# Tunneling-current-induced local excitonic luminescence in p-doped WSe$_2$ monolayers

Ricardo Javier Peña Román[*a], Yves Auad[a], Lucas Grasso[a], Fernando Alvarez[a], Ingrid David Barcelos[b] and Luiz Fernando Zagonel[*a].

[a]Applied Physics Department, "Gleb Wataghin" Institute of Physics, University of Campinas – UNICAMP, 13083-859 Campinas, SP, Brazil.
[b]Brazilian Synchrotron Light Laboratory (LNLS), Brazilian Center for Research in Energy and Materials (CNPEM), 13083-970 Campinas, SP, Brazil.
E-mail: rikrdopr@ifi.unicamp.br, zagonel@ifi.unicamp.br

## Abstract

We have studied the excitonic properties of exfoliated tungsten diselenide (WSe$_2$) monolayers transferred to gold substrates using the tunneling current in a Scanning Tunneling Microscope (STM) operated in air to excite the light emission locally. In obtained spectra, emission energies are independent of the applied bias voltage and resemble photoluminescence (PL) results, indicating that, in both cases, the light emission is due to neutral and charged exciton recombination. Interestingly, the electron injection rate, that is, the tunneling current, can be used to control the ratio of charged to neutral exciton emission. The obtained quantum yield in the transition metal dichalcogenide (TMD) is ~5 × 10$^{-7}$ photons per electron. The proposed excitation mechanism is the direct injection of carriers into the conduction band. The monolayer WSe$_2$ presents bright and dark defects spotted by STM images performed under UHV. STS confirms the sample as p-doped, possibly as a net result of the observed defects. The presence of an interfacial water layer decouples the monolayer from the gold support and allows excitonic emission from the WSe$_2$ monolayer. The creation of a water layer is an inherent feature of the sample transferring process due to the ubiquitous air moisture. Consequently, vacuum thermal annealing, which removes the water layer, quenches excitonic luminescence from the TMD. The tunneling current can locally displace water molecules leading to excitonic emission quenching and to plasmonic emission due to the gold substrate. The present findings extend the use and the understanding of STM induced light emission (STM-LE) on semiconducting TMDs to probe exciton emission and dynamics with high spatial resolution.

## 1. Introduction

Group-VI transition metal dichalcogenides (TMDCs or TMDs), like MoS$_2$ or WSe$_2$, are two dimensional (2D) semiconductors that display a transition from indirect band gap to direct band gap when thinned to a monolayer.[1] These materials have interesting properties like high quantum efficiency (QE), and high exciton binding energy and are flexible.[2–6] Moreover, some of their features can be engineered by defects which may, for instance, enhance their QE and act as single-photon emitters.[7–10] Such properties have triggered considerable interest for applications in several areas, including light-emitting devices and photovoltaic panels, among many others.[11–14] However, the exciton dynamics in the nanometer scale is far from been understood, despite the profound effect it has on the optical and electronic properties of such materials. The reports mentioned above lack nanometer-scale information on the exciton dynamics due to the resolution limit of widespread optical techniques, like photoluminescence (PL).

Recently, Pommier et al.[4] successfully applied Scanning Tunneling Microscopy Induced Light Emission (STM-LE, also referred to as STM induced Luminescence, STML) to study MoSe$_2$. They observed excitonic emission and confirm the diffusion of excitons at room temperature up to 2 micrometers. Their study was performed in MoSe$_2$ mechanically transferred to an ITO substrate. Previous STM-LE studies of gold or graphene supported TMDs could not observe excitonic emission even in quasi-freestanding MoS$_2$ (instead, reported light emission is due to radiative plasmonic modes).[15,16] Moreover, in the case of metallic supports, the Purcell effect could have even enhanced the light emission if it is possible to void quenching by fast charge transfer to the metal.[17,18] Indeed, for TMD monolayers, similarly to STM-LE performed on molecules, understanding the possibly quenching effect of the supporting substrate is crucial.[17,19–21] However, intriguingly, metallic substrates have been observed both to quench and to enhance luminescence from TMDs.[18,22–24] Also, recently both tip-enhanced PL and subsequent PL quenching for short tip–sample distance were observed for WS$_2$ and WSe$_2$ monolayers, including on gold substrates.[25–27] Finally, very few results of STM-LE applied to study the TMDs have been reported so far, even if STM-LE is an undeniably powerful technique to investigate the exciton dynamics and the optoelectronic response of 2D semiconductors together with their electronic structure and morphology at the nanometer scale.[28] Furthermore, the enhancement or quenching effects in the light emission of TMDs due to metallic substrates are ambiguous in the literature.

Here, we report the first observation of excitonic emission from a TMD on a metallic substrate using electron tunneling current as a nanoscale excitation source. We studied exfoliated WSe$_2$ monolayers transferred to gold thin film substrates and locally excited them using the tunneling current in an STM. STM-LE and PL spectra present the typical spectroscopic characteristic associated with the radiative recombination of bright A-neutral excitons and A-trions (charged excitons). The charge injection rate, adjusted by setting the tunneling current setpoint, controls the ratio of light emission due to trions and to neutral excitons. STM images and Scanning Tunneling Spectroscopy (STS) evidenced the presence of intrinsic struc- tural defects and confirmed the p-type doping of the WSe$_2$. The STM-LE excitation mechanism proposed is the injection of negative charges into the conduction band of the semi- conductor. Injected negative charges form neutral and charged excitons by Coulomb interaction with the positive charges in the p-doped sample. We propose the presence of a thin water decoupling layer to explain the absence of (excitonic light emission) quenching (due to the metallic substrate). Without quenching, excitonic STM-LE and PL become possible. The luminescence can be quenched on the whole flake by thermal annealing under ultra-high vacuum conditions or very locally removing the interfacial water using the tunneling current at the STM tip position. The results presented here give a clearer understanding of the excitonic light emission in STM-LE applied to TMDs and the quenching effect in 2D semiconduct- ing materials supported by metallic substrates.

## 2. Experimental details

The STM-LE experiments were performed using an STM (RHK Technology Inc.), where a light collector device has been installed. In the setup, both STM-LE and *in situ* PL can be per- formed, the latter using a 532 nm laser diode operated below 100 μW. Spectra are recorded using an imaging spectrometer coupled to a cooled CCD camera. *In situ* PL experiments were performed with the tip retracted but otherwise on the same sample location. STM-LE signal can also be recorded panchro- matically using a photomultiplier tube (PMT). STM images were acquired in the constant current-mode using a virtually grounded tungsten tip and a bias voltage applied to the sample. Tungsten tips were prepared by electrochemical etching with a NaOH solution and using a circuit for quickly stop the etching process. Tips were replaced frequently (at least every three days) to avoid thick tungsten oxide layers. Differential conductance was measured by STS using the lock-in technique with a modulation voltage of 40 mV and a fre- quency of 650 Hz.[29]

Monolayers were obtained from the same WSe$_2$ crystal purchased from HQ Graphene (p-type, having typical charge carrier densities of ∼$10^{15}$ cm$^{-3}$ at room temperature and a hole density of $p$ ∼ 8 × $10^{12}$ cm$^{-2}$). Flakes of WSe$_2$ were exfo- liated on commercial PDMS films (Gel-Film® PF-40-X4 sold by Gel-Pak) using blue tape. The PDMS stamp with WSe$_2$ was placed on a transparent quartz plate and upon slowly bringing WSe$_2$ in contact at room temperature with a gold thin film substrate (∼100 nm), deposited by sputtering on silicon. The ensemble was heated to ∼65 °C for two minutes using a Peltier module kept underneath the Si/Au substrate. After allowing the ensemble to cool down, the PDMS stamp was slowly detached, leaving behind the WSe$_2$ flake transferred on top. This transfer process was performed in a cleanroom at (23 ± 1) °C and a humidity of (55 ± 5)%. STM imaging and STS were performed under UHV (base pressure 5 × $10^{-8}$ Pa) in samples that undergone annealing also under UHV. STM-LE was per- formed in air in as transferred samples at room temperature, (23 ± 1) °C, and in a controlled humidity of (35 ± 5)%. In total, six large monolayers similar to Fig. 1(a) were prepared follow- ing this recipe and were studied. The presented results show a compilation that is entirely consistent with all acquired data. Raman spectroscopy and *ex situ* PL were performed out-side the STM in a Horiba MicroRaman system using a 532 nm laser at 10 μW.

## 3. Results and discussion

### 3.1. Monolayer WSe$_2$ on gold decoupled by a water layer

Fig. 1(a) and (b) show light optical and Atomic Force Microscopy (AFM) images of a typical WSe$_2$ flake, respectively. The gold substrate gives sufficient visual contrast to assist in identifying monolayers by optical microscopy. Raman and PL of the as transferred samples are similar to those observed in monolayers transferred to silicon dioxide substrates (see Fig. S1 in ESI†).[30–33] The height of the monolayer WSe$_2$ above the substrate is about 2.4 nm and, considering that the thick- ness of the monolayer is about 0.9 nm,[34] the height profile shown in Fig. 1(b) indicates the presence of the expected water layer under the 2D material transferred to a substrate in ambient conditions (see also Fig. S2†).[35–42] A detailed study of the presence of interfacial water and its effects on exfoliated MoS$_2$ was recently published by S. Palleschi *et al.*[43] In their study, a nano-confined water layer at the interface (TMD-sub- strate) has been probed by AFM in as transferred and annealed samples. Similarly, we have performed AFM measurements in as transferred and in annealed samples, and, as shown in Fig. S2 of the ESI,† the height profile of the annealed sample gives 0.9 nm, the expected thickness for a single WSe$_2$ mono- layer, typically observed in as-grown monolayers.[34,44,45] For WSe$_2$ flakes, water trapped between the exfoliated WSe$_2$ and the substrate can also lead to bottom flake surface oxidation.[46]

In ESI, Fig. S3† illustrates the STM tip approach process on he flake of interest for STM, STS, and STM-LE measurements. As prepared, the samples had atmospheric residues and trans- fer polymer residues on the surface. Such

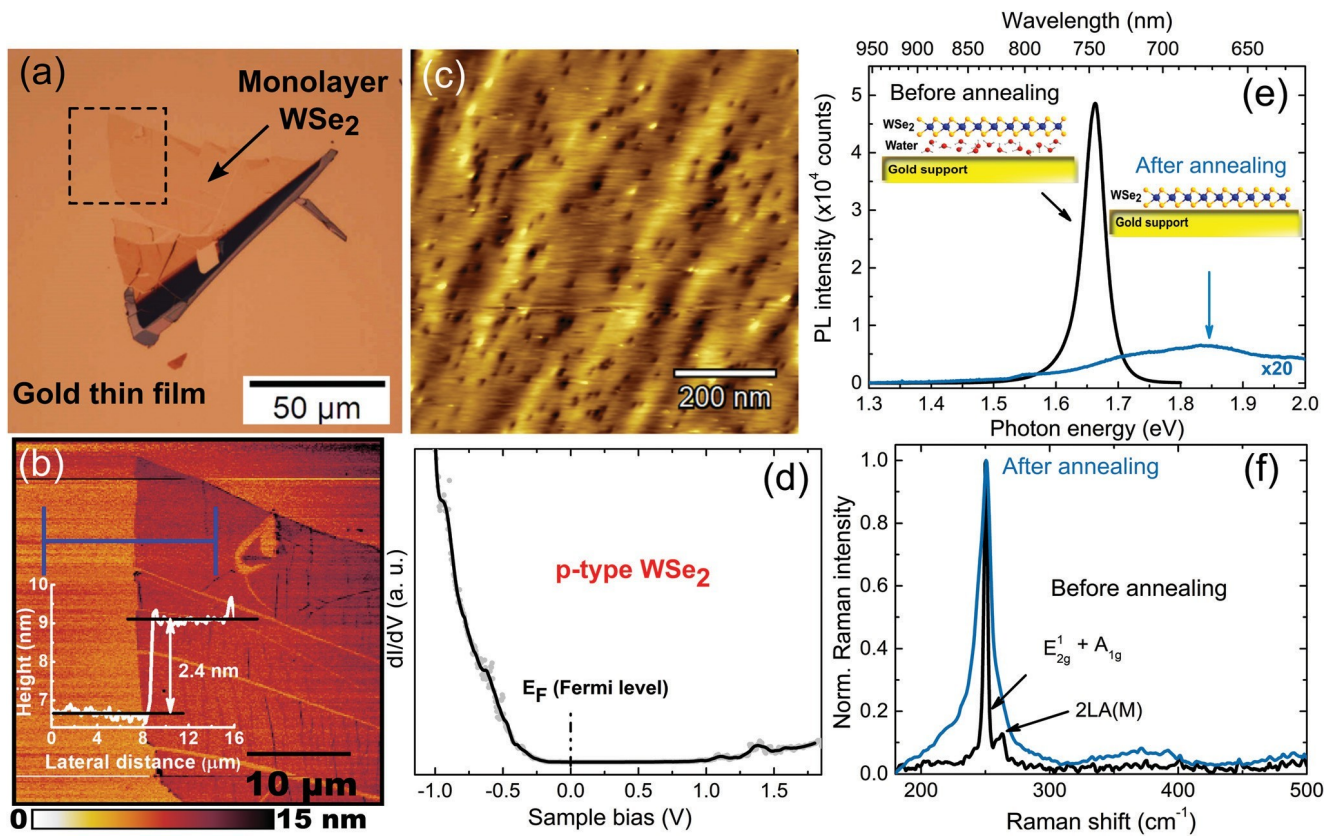

Fig. 1 (a) Light optical microscopy image of a typical WSe$_2$ flake. (b) AFM image of a region indicated in (a) by the dashed square. The average step height is 2.4 nm, as shown in the insert. (c) STM image obtained at room temperature and in UHV using a sample bias of 1.0 V and tunneling current of 65 pA. (d) d$I$/d$V$ spectrum showing the p-doped nature of the WSe$_2$ sample. (e) PL and (f) Raman spectra before and after the UHV thermal annealing on the same monolayer WSe$_2$. The inserts in (e) illustrate the sample-substrate decoupling by the interfacial water layer.

residues make them unsuitable for STM imaging or STS, both in air and under UHV. Samples are cleaned by mild thermal annealing at 400 K for 12 hours under UHV, as performed in other studies.[47] Annealing under these conditions should not modify the sample, and we consider it to be still the pristine WSe$_2$ mono- layer only outgassed and cleaner after such process. After annealing, images as the one shown in Fig. 1(c) could be acquired, showing a surface undulation revealed as a height modulation. STM images recorded in TMDs samples fre- quently observe this type of surface morphology. It could be related to different sample substrate distances as proposed by I. Delač Marion *et al.*[47] or to the way the monolayer attaches to the (not atomically flat) gold substrate (see Fig. S4(a) in ESI†). Also, STM images show the presence of defects as bright and dark spots, similar to reported in the literature.[48–51] The appar- ent defect profile width and height/depth largely depend on the bias voltage,[50,51] and so does their visibility. As shown in Fig. S4,† the observed depth, ∼0.5 nm, and width, ∼5 nm, of the dark defects are roughly similar to (depth) or significantly larger (by a factor 2 to 3, width) than reported for MoS$_2$ and MoSe$_2$.[49,50] The defects observed here are intrinsic in the pris-tine sample,[48] since thermal annealing treatments at (or below) 400 K in TMDs should not induce defect creation but should only clean the surface from atmospheric contaminants.[43,52–54] The density of dark defects is about $4 \times 10^{10}$ defects per cm$^2$, and of bright defects is $1 \times 10^{10}$ defects per cm$^2$ (see Fig. S4 in ESI†). The order of magnitude of these intrinsic defects densities is consistent with reported results in exfoliated samples.[49,52,55] Previous STM/STS studies of point defects in TMD materials have shown that dark defects are related to transition metal defects (vacancies) inducing p-doping in the sample whereas bright defects are associated with chalcogen defects or adsorbate impurities with n-doping effect.[48–51,56,57] Therefore, since ∼80% of the defects are of the dark type, the nominal p-doped character of the sample can be associated with dark defects. However, the difference between nominal p doping and the observed defect density could indi- cate other sources of doping or the existence of more defects than those observed in Fig. 1(c).[50]

We measured the density of states (DOS) using STS. In the STS curve, presented in Fig. 1(d), the Fermi level is closer to the valence band edge, indicating the hole-doped (p-type) nature of the sample. This result agrees with the expected (nominal) doping type and with reported for WSe$_2$ growth on highly oriented pyrolytic graphite (HOPG).[51] Additionally, in Fig. 1(d), the low DOS in the conduction band could indicate band bending, such as has been perceived in ref. 48, 49, 58 and 59. The reported band gap in the literature for exfoliated WSe$_2$ monolayers at room temperature lays between 1.9 and 2.0 eV.[60,61] However, the estimation or determination of a band gap value from the spectrum in Fig. 1(d) might be inaccurate since band edges are not observed clearly. The d$I$/d$V$ curve may also be affected by the adhesion of the monolayer with the metallic substrate, such as found in systems such as MoS$_2$/Au(111),[15,62,63] WS$_2$/Ag(111) and WS$_2$/Au(111).[64] In such reports, samples are epitaxially grown on monocrystalline substrates and the interaction between the TMD and the metallic surface causes a hybridization of semiconductor states with substrate states leading to a band gap reduction. Also, it has been reported that this hybridization may produce other effects such as doping or metallization of the 2D semi-conductor. However, these effects are expected only in epitaxial samples with high interfacial quality, and not in samples prepared by mechanical exfoliation methods.[64] The latter and the fact that the metallic substrates used in this work are not atomically flat could help explaining why, after all, to some extent, the STS data in Fig. 1(d) might be used as confirmation of the doping type. Finally, it is important to note that the STS in Fig. 1(d) refer to the monolayer WSe$_2$ absorbed on the gold substrate and it is a different scenario with respect to the water decoupled monolayer.

Fig. 1(e) and (f) show the (*ex situ*) PL and Raman spectra measured in a monolayer WSe$_2$ in air and at room temperature. These results were obtained before and after the UHV thermal annealing treatment. Before annealing, PL shows a sharp excitonic peak centered at 1.66 eV (747 nm), and the Raman spectrum shows the characteristic splitting of 12 cm$^{-1}$ between the degenerate $E^1_{2g}$ + A$_{1g}$ and the 2LA(M) Raman modes around 250 cm$^{-1}$. One must note that PL experiments on TMDs are generally performed on samples grown on (or mechanically transferred to) isolating substrates like silicon dioxide.[9,38,60,65] Studies of TMDs on metallic substrates report luminescence enhancement or quenching, depending on the growth or transfer method.[15,18,22,66,67] However, it is known that mechanically transferred 2D material encapsulates thin airborne water (moisture) layers.[36,38,43,46] Such water layer could act as a spacer (also called barrier or buffer layer)[68] to decouple them and avoid the quenching by the metallic substrate, such as is illustrated in the inserts in Fig. 1(e). This decoupling layer is very similar to the needed spacer to observe STM luminescence of molecules in conducting substrates: a spacer that decouples to prevent quenching but allows electric contact within a reasonable compromise.[19–21] We consider the existence of a water layer beneath the monolayer WSe$_2$, as evidenced by AFM results in Fig. 1(b) and S2† after the mechanical transfer to the gold substrate. This layer acts as a dielectric insulating layer, and it is responsible for the effective semiconductor–metal (sample–substrate) decoupling and the observed PL and Raman spectra from Fig. 1(e) and (f).

After annealing, the step height, shown in Fig. 1(b) and S2(f),† reduces from 2.3 nm to 0.9 nm, as shown in Fig. S2(i),† indicating the removal of the water layer by the thermal annealing. That is, the step height after annealing is the mono- layer thickness, suggesting the evaporation of the water that was under the monolayer. The drastic change in luminescence and Raman response after annealing also firms the vanished decoupling layer. After the annealing, the PL signal is severely quenched, and Raman peaks are broader (with respect to the Raman peaks in the as transferred samples) and the splitting between Raman peaks $E^1$ + A$_{1g}$ and 2LA(M) is gone. These results suggest that the annealing produced a relevant electronic and mechanical contact between the monolayer and the gold substrate. This is consistent with the STS results which show that the sample–substrate interaction produces an electronic hybridization. The Raman peak widths could be related to stress gradients due to the monolayer adhesion on the gold thin film surface (see Fig. S4(a)†). Curiously, water intercalation enhanced the PL of sapphire supported WS$_2$, indicating its relevance even for insolating substrates.[37] However, water moisture trapped under TMDs was also reported to induced doping and be deleterious for the QE (as seen in PL) of MoS$_2$ and WS$_2$ but had a negligible effect on WSe$_2$.[38]

### 3.2. Excitonic light emission induced by the tunneling current

Being demonstrated above that there will be a sample–substrate decoupling that avoids the luminescence quenching, we performed STM-LE experiments in air on as transferred WSe$_2$ monolayers without previous annealing. The insert in Fig. 2(a) shows the STM tip approached on top of a monolayer WSe$_2$, as observed to find the sample. Fig. 2(a) shows STM-LE and *in situ* PL spectra obtained sequentially from the same monolayer WSe$_2$. In STM-LE, the sample bias was 4.0 V with a current setpoint of 44 nA in a fixed tip position. Panchromatic detection of the STM-LE signal using a PMT gave a Quantum Yield (QY) of ∼ 5 × 10$^{-7}$ photons per electron (see Fig. S5 in ESI†), a similar value to the reported for MoSe$_2$.[4] As shown in Fig. S5,† using a PMT, it is also possible to observe fluctuations of the light emission as expected in air.[69] Fig. 2(b) (see also Fig. S6†) shows STM-LE spectra acquired with different sample bias, each with a fixed STM tip position on the sample surface at different regions inside an area of 100 nm × 100 nm. No light could be detected for bias voltages of 1.9 eV or below, in agreement with the reported band gap, 2.0 eV.[61]

As pointed out in ref. 4 and 19, the close resemblance between the PL and STM-LE spectra (Fig. 2(a)) and the constant emission energies for different bias voltages in STM-LE spectra (Fig. 2(b)) are evidence that the emission mechanism is the same: radiative recombination of bright A excitons through the direct band gap of the semiconductor,[70] see also Fig. S7.† As shown in Fig. S8,† STM-LE spectra as a function of the sample bias voltage show no systematic peak energy shift, thus discarding any plasmon related emission from the tunneling cavity. Also, plasmon emission is distinctively different

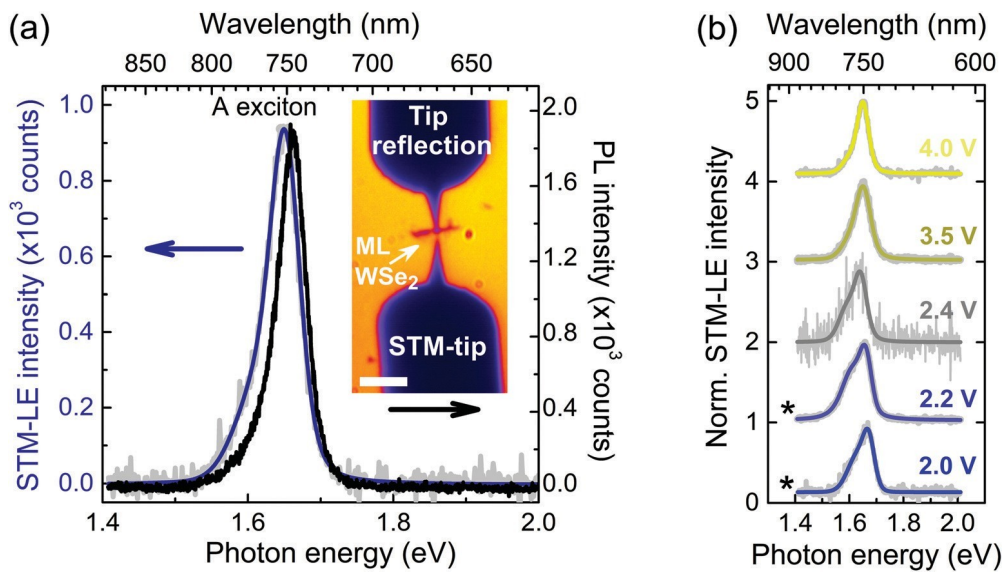

Fig. 2 (a) PL and STM-LE spectra of a monolayer of WSe$_2$. PL is excited with a laser (λ = 532 nm) when the tip is retracted. For STM-LE, the sample bias was 4.0 V, the current setpoint was 44 nA, and the integration time was 100 s. Insert: Zoom-lens optical micrograph showing the STM tip on top of the same flake shown in Fig. 1. The scale bar has 100 μm. (b) STM-LE for different sample bias voltages and different tunneling currents. The symbol * indicates those spectra taken using 25 nA as tunneling current, whereas the current was 44 nA for the others. In the spectra, the gray curves represent the raw data, and the solid colored lines are the result of curve fittings.

from WSe$_2$ excitonic emission, as discussed below and shown in Fig. S9.† The small peak position variations shown in Fig. S8† are due to the different acquisition positions for each spectrum. Indeed, each spectrum acquisition is performed in a new tip position to avoid changing the sample with the high tunneling current, as discussed in detail in ref. 4. Such peak energy variation happens because each different region of the WSe$_2$ layer is possibly subject to different stress due to the substrate roughness and possibly contains a slightly different density of defects.

In TMDs, the difference between the electronic band gap energy and the emitted photon energy is due to the exciton binding energy, 0.37 eV, that prevents the band-to-band radiative recombination and leads to excitonic light emission with smaller energy than the band gap even at room temperature.[61] Additionally, neutral excitons and charged excitons (trions) dominate the optical properties of TMDs.[60] STM-LE spectra in Fig. 2(a) is a convolution of the radiative recombination of bright A-neutral excitons at 1.65 eV (751 nm) and A-trion at 1.62 eV (765 nm). Fig. S7† shows fitted spectra in close agreement with recent literature.[71–73] In the PL spectra, the neutral exciton and trion emissions are observed at 1.66 eV (747 nm) and 1.63 eV (761 nm), respectively. The difference between the neutral exciton and trions peaks gives a trion binding energy of 30 meV. As mentioned previously, the neutral or charged exciton emission energies are independent of the bias voltage or tunneling current (see Fig. S8†). Fig. S8† indicates that the observed light emissions are related to what is happening inside the WSe$_2$ monolayer and do not depend on the proximity of the tip (tunneling cavity size) or the electric field under the tip.

### 3.3. Trion emission control using the tunneling current

The control and manipulation of neutral excitons and trions are crucial for improving and tuning the performance of TMDs based devices. Several approaches have been used with this propose in the last years, such as chemical doping,[74] gate doping,[75] photoexcitation,[76] and charges transfer in TMDs heterostructures.[77] More recently, He et al. have reported the nanoscale control of negative trions using a plasmonic picocavity in WS$_2$ monolayers.[25] In STM-LE, the charge carrier injection rate given by the tunneling current can control A trion emission. Fig. 3(a) shows STM-LE spectra acquired for different tunneling currents indicating this systematic change of the light emission related to trions for currents ranging from 25 nA to 90 nA. Below 25 nA, no exploitable spectrum could be acquired. As in other cases, each spectrum was acquired in a different (fresh) region to avoid any possible sample modification or cumulative effects. Also, optical microscopy before and after these measurements spotted no sign of STM related sample damage.

As shown in Fig. 3(b), which summarizes the data of Fig. 3(a) and other spectra not shown, the area of the trion peak, $A_T$, divided by the area of the neutral exciton, $A_0$, (area ratio $A_T/A_0$) scales linearly with the tunneling current. Interestingly, the low laser power PL spectrum in Fig. 2(a), also represented in Fig. 3(b) as the point at 0 nA, shows the ratio 0.5 when both electrons and holes are created with relatively low density (laser power below 100 μW in a spot with ~2 microns in diameter). In this case, trion formation relates to the sample doping. In STM-LE, with single carrier type (electron) injection, a 0.5 AT/A0 ratio would correspond roughly to a tunneling current of 20 nA (that is below our detection limit).

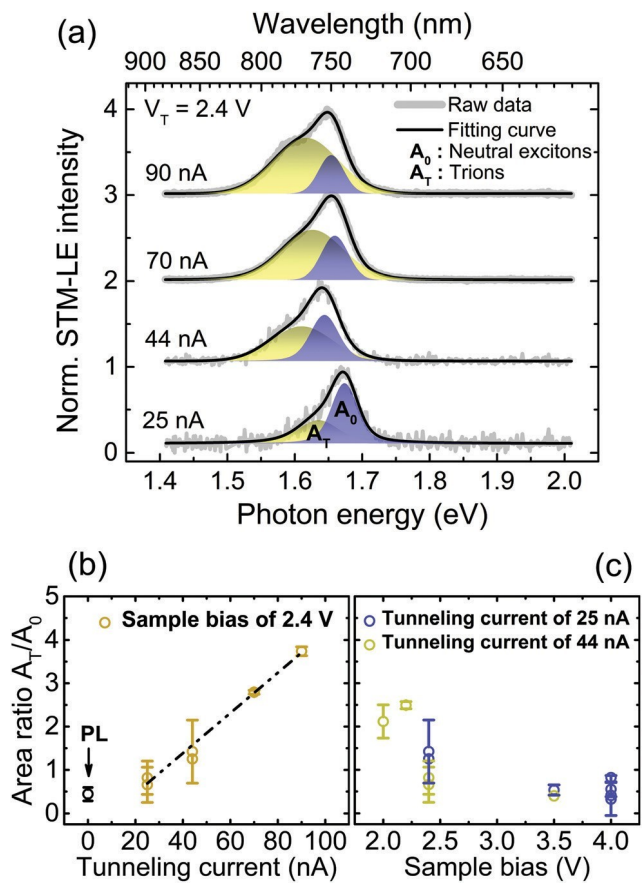

Fig. 3 (a) STM-LE spectra for 2.4 V (sample bias) shows the effect of different tunneling currents. Higher currents enhance trion emission. (b) and (c) show the dependence of the ratio of trion to neutral exciton emission as a function of the tunneling current and sample bias, respectively. In (b), the ratio for the PL spectrum is also indicated.

20 nA could be considered as the maximum charge injection rate that does not increase trion formation by an excess of injected electrons. On the other limit, at 90 nA, the trion emission is nearly four times greater than the emission of neutral excitons.

Considering the exciton lifetime that ranges from 0.4 ns to 4 ns in WSe$_2$ monolayers at room temperature[78,79] and the time interval between consecutive injected electrons, that is, for instance, 6.4 ps in average for a tunneling current of 25 nA, the tunneling current should affect the trion formation. That is, at low currents, each injected electron would have time to find a hole and form an exciton and latter possibly form a trion with another hole from the sample. At high currents, there would be an increased probability that an injected electron would find an already formed exciton before it had time do decay or to diffuse away. Using the diffusion coefficient of excitons measured by S. Mouri *et al.* in ref. 78, $\sim(2.2 \pm 1.1)$ cm$^2$ s$^{-1}$, and the time interval between consecutive injecting electrons (supposing one electron arrives at a time in regular intervals), an exciton created under the tip could diffuse less than $10^2$ nm before the next one arrives at 25 nA and half of this distance for 90 nA. Additionally, at least for high currents, one can consider that the observed trions are most likely negatively charged since the overwhelming electron injection forms them into the p-doped material. A more quantitative analysis, however, would require details that go beyond the scope of this work. In contrast to reported in ref. 25, where trions are controlled in a light excited picocavity, here we show that, in STM-LE, trions could be controlled by the tunneling current at the nanoscale. Dynamical effects due to the charge injection rate on light emission have also been explored in other semi- conducting systems using STM-LE and Cathodoluminescence in Scanning Transmission Electron Microscope (STEM-CL).[80,81]

Whereas the effect of the tunneling current on the charged to neutral exciton ratio is about fivefold in the available tunneling current range and can be interpreted as an accumulation of injected negative charges, changing the sample bias also presents a similar effect, as shown in Fig. 3(c). Indeed, for constant current and increasing sample bias, the ratio changes from ∼2.5 to ∼0.5 (the ratio decreases for increasing bias). This effect is not expected in the simple view of charge injection. However, in STM, changing the sample bias changes more than the end state of elastically tunneled electrons. One could expect different ratios of elastic to inelastic tunneling, among other effects. In this sense, one must consider that the bias voltage has a broader impact on the experiment than the tunnel current, and a comparison between results obtained under different bias voltages requires care. It is worthy to note that the trion to neutral exciton ration increases for decreasing tip distance in both situations in Fig. 3(b) and (c). Nonetheless, as shown in Fig. S8,† the bias voltage and the tunneling current have no significant effect on the trion and neutral exciton emission energies, indicating the absence of energy shifts due to the tip electric field.

### 3.4. Local excitonic emission quenching

We also observed that the excitonic light emission quenching due to the gold substrate could be produced locally by the tunneling current. Fig. 4 shows STM-LE spectra measured sequentially in the same point of the monolayer just after moving the STM tip to a new region and keeping the tip position fixed during the spectra integration time (100 s per spectrum). The sample bias was 4.0 V and the tunneling current setpoint 44 nA. We note that the excitonic emission at 1.65 eV progressively vanishes, and a broad plasmonic emission appears. This plasmon emission is interpreted as a cavity mode between the tip and the gold substrate surface with the TMD in the middle, similarly as previously observed for MoS$_2$.[15] Other tunneling parameters can also lead to plasmon emission after the excitonic light emission from the WSe2 is quenched (see Fig. S9†). Particularly, approaching the tip to create a brief contact between sample and tip (a tip reconstruction process called 'tip pulse') quickly quenches excitonic emissions and triggers plasmonic emission. The tip and the tunneling current causes the progressive change from excitonic to plasmonic emission by the local removal of water molecules that move away from the tip position, as already observed in graphene transferred on mica[39] and MoS₂/HOPG samples.[40] The direct sample–sub-

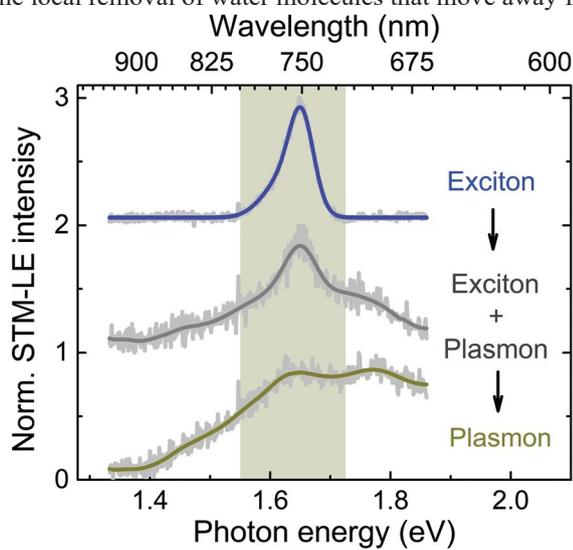

Fig. 4 STM-LE spectra taken sequentially in the same sample region with the tip position fixed (44 nA, 4.0 V, and 100 s). The highlighted region shows the position of excitonic emissions.

strate contact when the water under the monolayer is removed quenches exciton emission locally by creating fast non-radiative recombination paths. A transition from excitonic to plasmonic emission has also been systematically obtained in

STM-LE studies on molecular systems by varying the effectiveness of the decoupling between the optically active molecule and the conducting substrate.[19,28] It is interesting to note also that plasmon emission could be readily obtained using the parameter given in Fig. 4 in some regions. However, in other regions, pure excitonic emission would persist for longer times, possibly due to the gold support topography. Also, using lower sample bias voltages, excitonic emission is observed for longer times with no significant plasmon emission. Finally, sample damage (defect creation) can be ruled out due to the absence of luminescence changes when using high current or high bias (see Fig. 2 and 3) and since plasmon emission appears when excitonic emission vanishes which indicate the collapse of the monolayer into the substrate (following water removal).

### 3.5. Excitation mechanism due to the tunneling current

In Fig. 5, we summarize a proposition of the light emission and excitation mechanisms in STM-LE in monolayer $WSe_2$ together with a scheme of the experiment and band diagrams. First, the emission mechanism, as discussed previously, is the neutral exciton and trion recombination, as observed in PL. The sample–substrate decoupling due to the substrate–borne water moisture is fundamental to avoid luminescence quenching by the gold substrate. Fig. 5(a) illustrates the experiment and the emission mechanism.

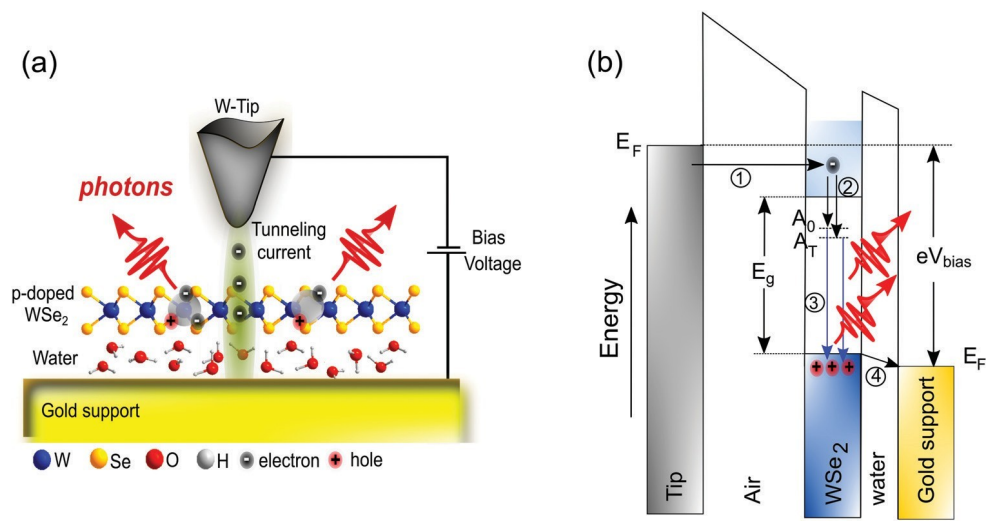

Fig. 5 (a) Scheme of the experiment showing the presence of the decoupling water layer. (b) Proposed STM-LE excitation mechanism. Elastically tunneling electrons are injected as hot carriers in the $WSe_2$ conduction band (1), after thermalization, the injected carrier binds into an excitonic state (2) then photons are emitted by exciton recombination (3), and, finally, electrons tunnel from the monolayer to the gold to keep the monolayer neutral (4). The sample bias voltage is larger than the electronic band gap.

Unlike PL, however, the excitation in STM-LE happens in the nanoscale, and the excitation mechanism is not the creation of electron–hole pairs. In this work, we propose the direct electron injection in the conduction band by the STM tip as the excitation mechanism. Subsequently, injected electrons form excitons and trions with the holes already present in the monolayer. The following results support this excitation mechanism: (i) the STM/STS results evidenced the presence of defects and p-type doping (in agreement with the nominal sample doping), which imply that positively charged carriers (holes) are present in the sample and, by injecting electrons, exciton formation is possible; (ii) no STM-LE signal was observed with negative biases, in connection to the p-doped nature of the sample, which also rules out the possibility of plasmon mediated emission;[82–84] (iii) excitonic light emission was only observed with positive bias voltages at or above 2.0 V, which is expected for electron injection luminescence mechanisms, where biases higher than the electronic band gap are necessary to excite the light emission.[19,85,86]

In Fig. 5(b), a band diagram summarizes such excitation mechanism. First, the electrons tunnel elastically into the conduction band of the 2D semiconductor. Then, they bound with existing holes into excitonic states. The substrate–borne water moisture acts as a dielectric barrier to decouple the monolayer and the gold substrate (support), preventing non-radiative recombinations due to substrate effect (quenching). Next, the recombination of neutral excitons and trions gives rise to the light emission. The possible band banding in the monolayer, if present, does not affect excitonic emission energies, as shown in Fig. 2(b) and S8.† Finally, an electron tunnels from the mono- layer to the substrate to neutralize the monolayer. Regarding the role of the water layer, one might consider its effects go beyond that of mere decoupling the monolayer $WSe_2$ and making it behave as if it were free-standing. Additionally, TMDs on top of trapped water layers have much smaller PL efficiency than truly suspended ones.[38] Indeed, the water layer might change the TMD quantum efficiency and light emission by chemical doping,[38] similarly as observed for aromatic solvent.[87]

Generally speaking, one must also note that, in STM-LE, only one carrier polarity is created (injected). Hence, exciton formation depends on the sample doping for the other carrier, while, in PL, both carriers are created inside the sample. Another striking contrast between these techniques is that, in STM-LE, the excitation is exceptionally local, certainly in the sub-nanometer range. While, in PL, the excitation region spams over at least nearly one squared micrometer. Also, it is important to note that the light detection region in STM is different from that in PL or in a Scanning Near-Field Optical Microscope (SNOM) in which both excitation and detection are local and similar.[28] In STM-LE, similarly to CL-STEM, the excitation is local (local excitation, possibly below one squared nanometre), but the region from which emitted light is detected is much larger (global detection).[88,89] This means that one knows where the excitation was performed (tip or electron probe position) but does not know the location of the emitted light. In the context of the local quenching due to the point contact between the monolayer $WSe_2$ and the gold substrate, global light detection means that all injected carriers are lost into the gold substrate before any diffusion in the TMD could take place. This is expected since the electrons in the $WSe_2$ conduction band should recombine non-radiatively through the metal very quickly (quench effect), and the vast majority of the injected carrier would follow this recombination path since the contact is precisely at the tip position.

## 4. Conclusions

We observed excitonic light emission from $WSe_2$ monolayers excited by the tunneling current within an STM in ambient conditions, using a non-plasmonic (tungsten) tip and on a metallic support. The luminescence excited by the tunneling current resembles that excited by light (PL) and is attributed to radiative recombination of bright A-neutral excitons and A-trions. The ratio of trions to neutral excitons emission can be controlled and manipulated using the tunneling current, but the emission energies remain unaffected by tunneling parameters. The gold support where the monolayers were transferred to did not quench the excitonic light emission from the monolayer due to an inherent interfacial water layer that effectively decouples them. Quenching could be triggered either by thermal annealing under vacuum, as confirmed by PL and Raman, or locally using the tip and tunneling current, as observed by STM-LE. Both processes remove the underneath interfacial water layer, either globally or locally, respectively. Without the water layer, plasmonic emission could be observed by STM-LE due to the gold substrate. These results contribute to the understanding of the STM-LE use in the investigations of light emission from TMDs and of the quench- ing effect due to metallic supports.

## Conflicts of interest

The authors have no conflicts of interest to declare.

## Acknowledgements

This work was supported by the Fundação de Amparo à Pesquisa do Estado de São Paulo (FAPESP) Projects 14/23399-9 and 18/08543-7. I. D. B. acknowledges the financial support from the Brazilian Nanocarbon Institute of Science and Technology (INCT/Nanocarbono) and Brazilian Synchrotron Light Laboratory (LNLS). L. F. Z. thanks Dr. Eric Le Moal (ISMO) and Dr. Mathieu Kociak (LPS) for valuable and insightful discussions which were exceedingly helpful to this research.

# Electronic Supplementary Information

## Tunneling-current-induced local excitonic luminescence in p-doped WSe$_2$ monolayers

Ricardo Javier Peña Román,*[a] Yves Auad, [a] Lucas Grasso,[a] Fernando Alvarez,[a] Ingrid David Barcelos [b] and Luiz Fernando Zagonel **[a]

Figure S1: Pre-characterization of WSe$_2$ samples

Figure S2: AFM profiles before and after thermal annealing

Figure S3: Localization of exfoliated monolayer WSe$_2$ inside STM

Figure S4: STM images in UHV conditions and densities of point defects in monolayer WSe$_2$

Figure S5: STM-LE quantum yield and temporal stability of the STM-LE signal

Figure S6: Excitonic STM-LE spectra with different tunneling parameters

Figure S7: Neutral exciton and trion emissions in STM-LE and PL spectra

Figure S8: Trion and neutral exciton emission ratio

Figure S9: Exciton and plasmon emission with different tunneling parameters

\* rikrdopr@ifi.unicamp.br

\*\* zagonel@ifi.unicamp.br

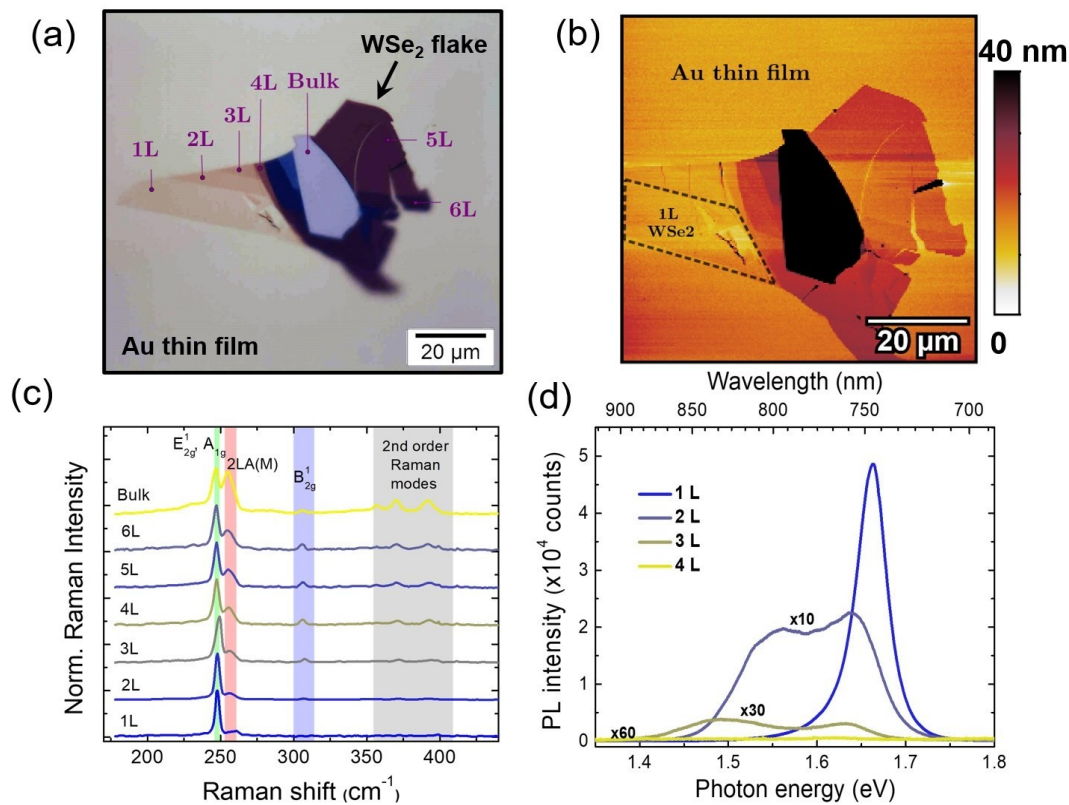

**Figure S1:** Pre-characterization of an exfoliated WSe$_2$ flake transferred to 100 nm Au thin film on Si substrate. (a) Optical microscopy image. The optical contrast in the gold substrate helps to identify regions with a different number of layers. (b) Atomic Force Microscopy (AFM) image showing regions with different heights and imperfections created in the flake during the transferring process. (c) Raman and (d) photoluminescence spectra at room temperature. The monolayer WSe$_2$ has been confirmed by the absence of the interlayer Raman mode ($B_{2g}^1$) around 300 cm$^{-1}$, as well as, by the strong PL peak at 1.65 eV. The Raman and PL spectra have the same characteristic of those observed in samples deposited on silicon dioxide substrates [1–4], indicating that in as-transferred samples there is no coupling between the TMD and the metallic substrate (quenching effects are not observed).

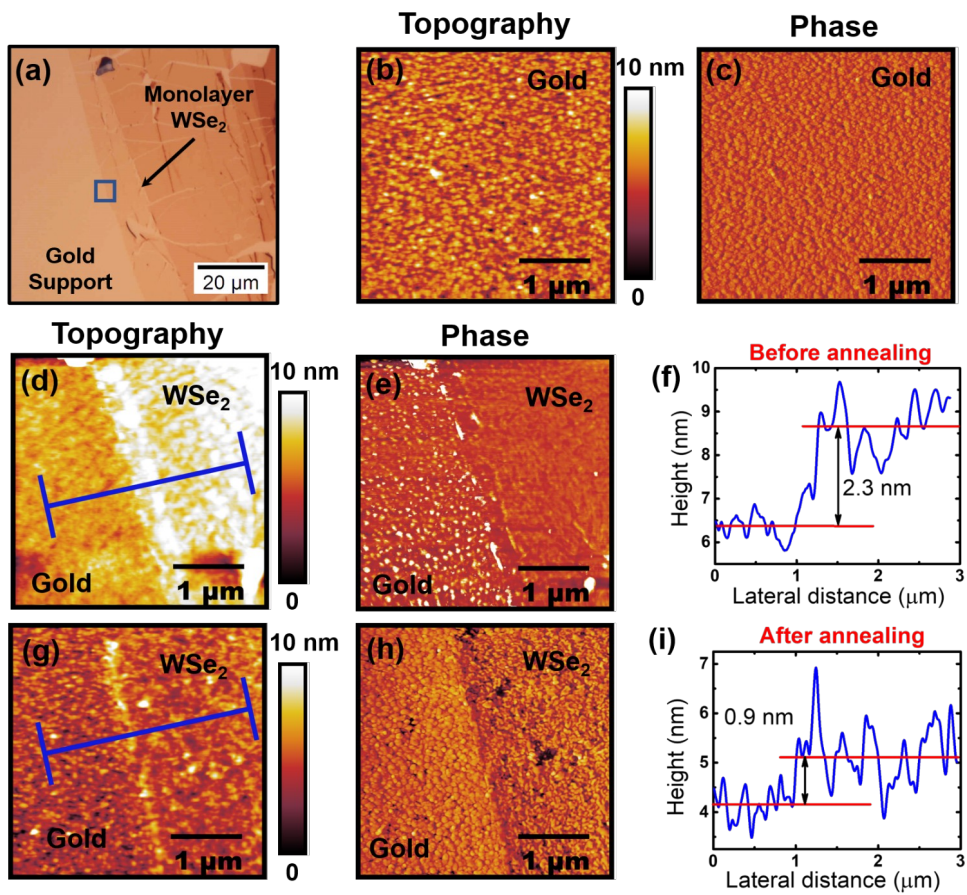

**Figure S2:** (a) Optical microscope image showing the Monolayer WSe$_2$ on top of the gold substrate. (b) Topography and (c) Phase AFM images of the gold substrate. (d) Topography and (e) Phase AFM images of the interface between the gold substrate and the monolayer WSe$_2$ before the 400 K thermal annealing in UHV for 12 hours. (f) AFM line profile obtained from (d) averaging a 3 mm long and 1 mm large area indicating a step height of 2.3 nm. (g) Topography and (h) Phase AFM images of the interface between the gold substrate and the monolayer WSe$_2$ after the 400 K thermal annealing in UHV for 12 hours. (i) The AFM line profile obtained from (g), averaging a 3 microns long and 1 mm large area, indicates ~0.9 nm of step height. Since the expected thickness of the monolayer WSe$_2$ is about 1 nm, we consider that before the thermal annealing in UHV there was a water layer under the TMD.

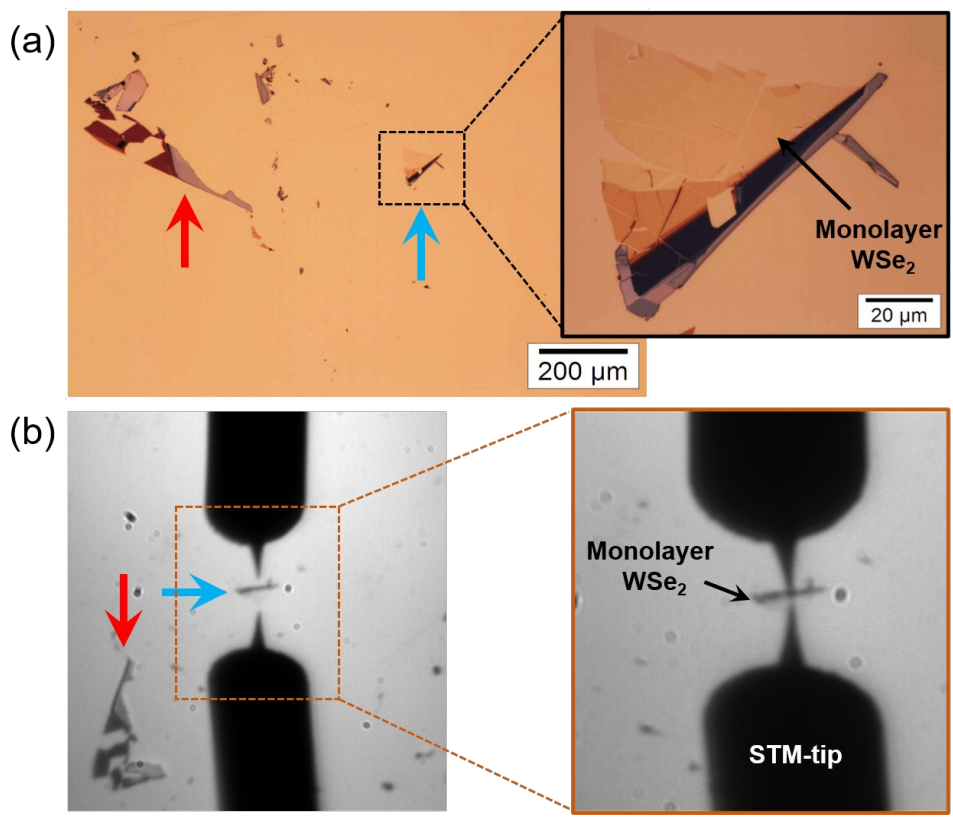

**Figure S3:** Localization of the flake of interest inside the STM microscope for the tip approach. (a) Optical microscopy image. The flake indicated by the red arrow is used as a reference point to localize the flake of interest with monolayer WSe$_2$ as indicated by the blue arrow. (b) By using a zoom lens, the monolayer WSe$_2$ is localized inside the STM, and the tip is approached. In the image, the STM tip and its reflection on the metallic substrate are observed. The tip wire diameter is 250 mm. This procedure is the same for STM measurements in UHV, and STM-LE in air.

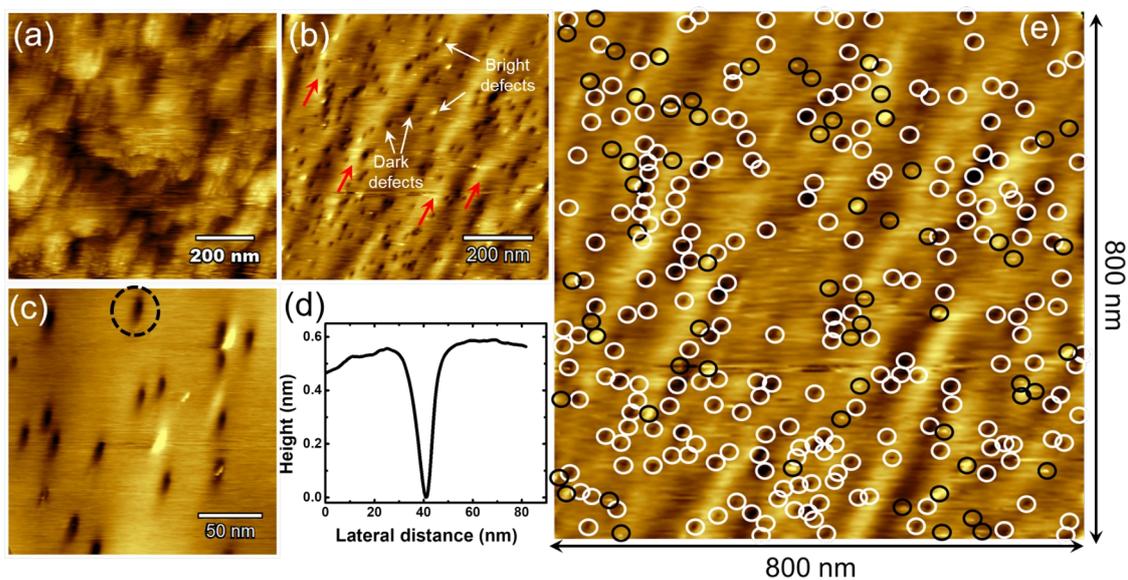

**Figure S4:** STM images obtained in UHV at room temperature in monolayer WSe$_2$/Au/Si annealed at 400 K. (a) STM-image of 1.0 ㎜ ´ 1.0 ㎜ of the Au thin film surface (3.5 V, 300 pA). (b) STM-image of 0.8 ㎜ ´ 0.8 ㎜ of the WSe$_2$ surface (1.0 V, 65 pA). A surface height undulation (red arrows) is observed in addition to some dark and bright point defects. (c) STM-image of 226 nm ´ 226 nm of the WSe$_2$ surface (1.0 V, 65 pA) showing dark defects and (d) the depth profile of one of them. (e) Quantification of the defects density in the image shown in (b). Dark and bright defects are indicated in white and black circles, respectively.

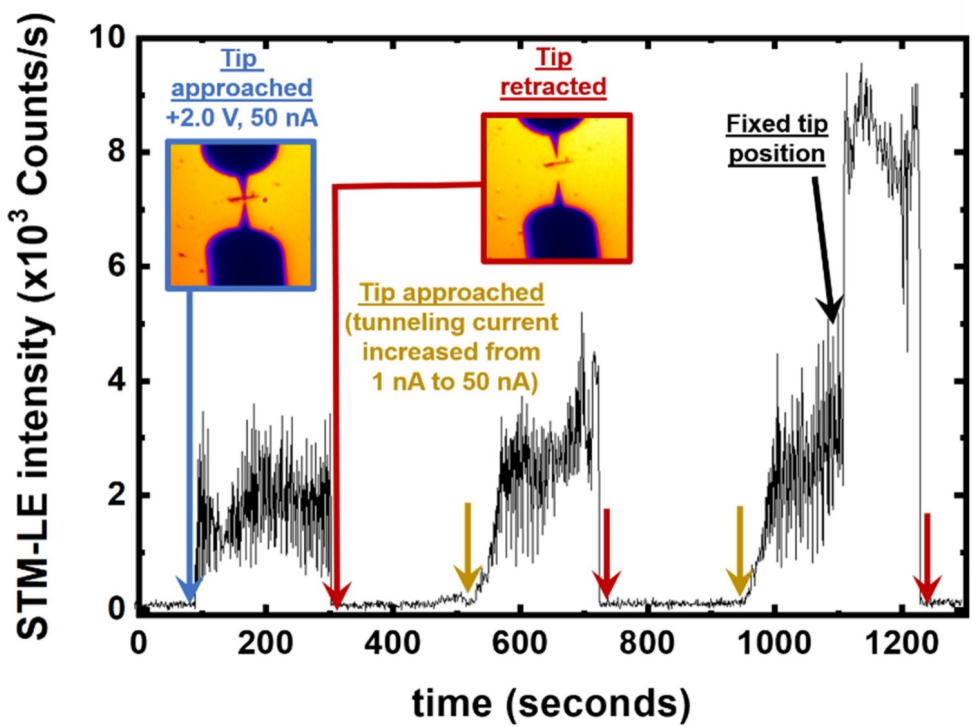

**Figure S5:** The STM-LE signal was recorded panchromatically as a function of time using a PMT to evaluate the quantum yield and temporal stability. Three cycles of tip approach and retraction are shown. Sample bias is set to 2.0V, and the tunneling current increased up to 50 nA. The results indicate the variation of the observed signal as well as the ease of getting the signal back after establishing a tunneling current. The signal variations are also observed due to the short PMT integration time: 0.3 seconds. Similar variations are typically observed in the tunnel current, as one could expect in air [5]. In the first 1100 s (~18 minutes) the signal was acquired scanning the tip on a surface area of 130 nm ´ 130 nm, after that the STM tip position was fixed, and signal of about $9\times10^3$ counts per second was obtained with 50 nA. This gives a Quantum Yield (QY) of ~5´$10^{-7}$ photons per electron, similar to observed in $MoSe_2$ mechanically transferred on ITO [6].

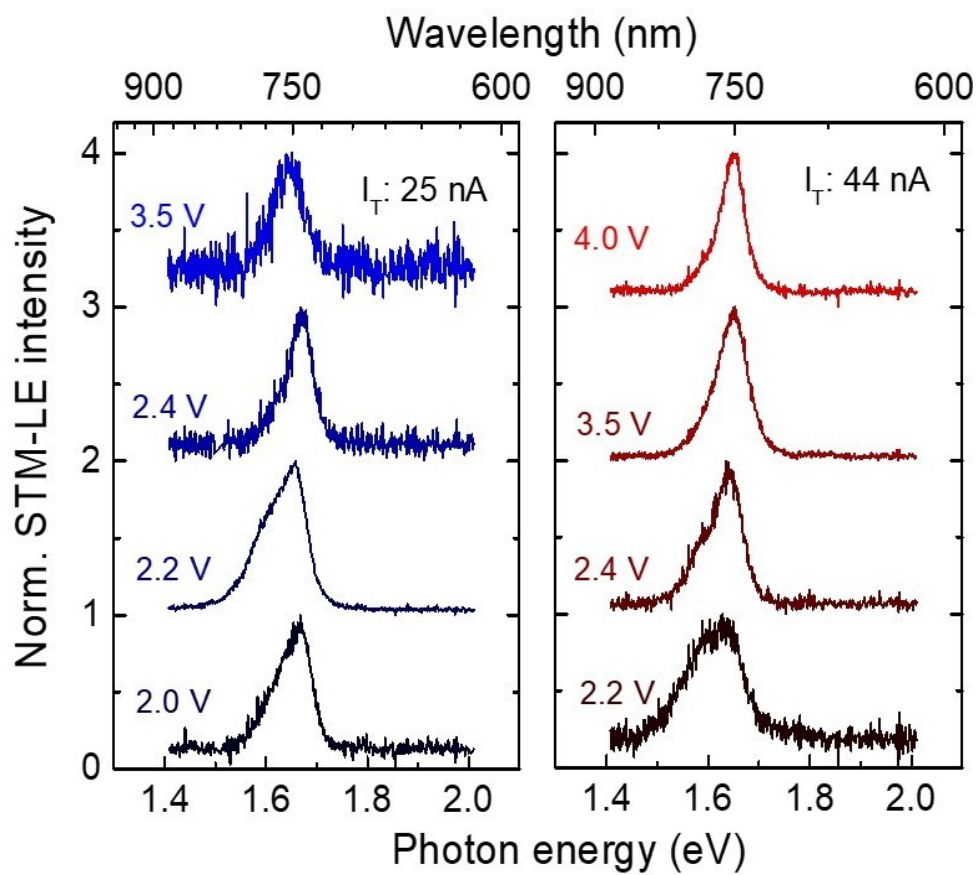

**Figure S6:** STM-LE spectra measured with different sample bias and tunneling currents of 25 nA (black to blue curves) and 44 nA (black to red curves), the integration time per spectrum was of 100 s. Some different peak shapes are observed due to the different intensities of the charged exciton (trion) and of the neutral excitons. See the decomposed spectra in Figure S7 and the ratio as a function of sample bias in Figure 3(c). The center of each emission remains roughly at the same energy, as shown in Figure S8. That is, no systematic energy shift is observed in the STM-LE peak as a function of the sample bias. The same spectral features are observed both at 25 nA and at 44 nA.

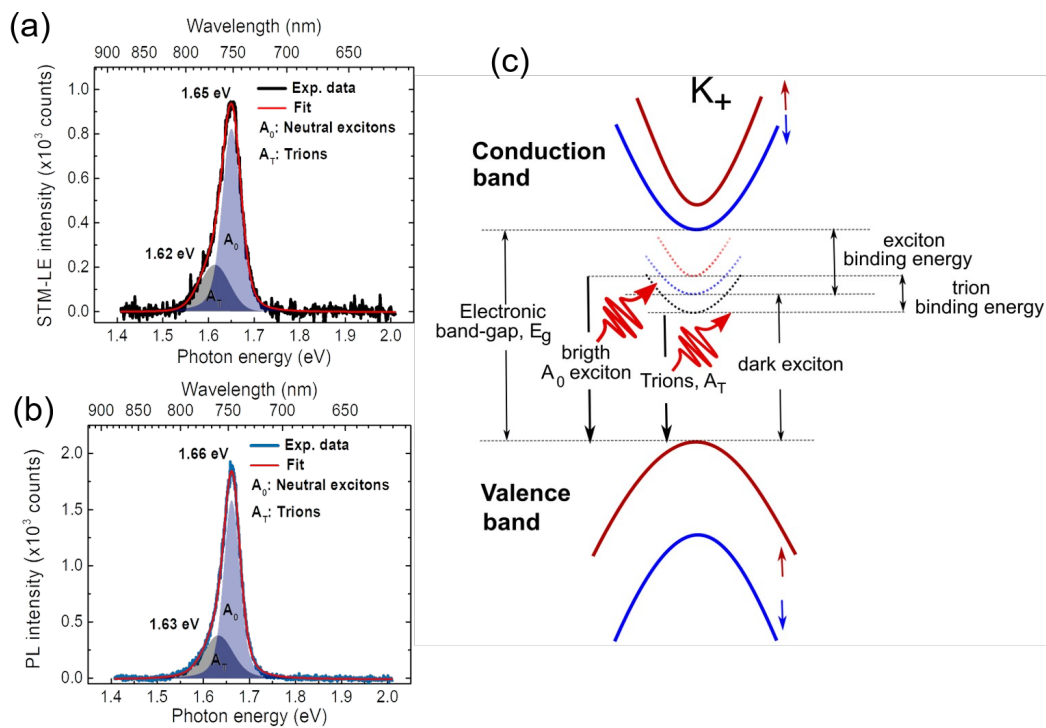

**Figure S7:** (a) STM-LE (4.0 V, 44 nA) and (b) PL spectra of monolayer $WSe_2$ fitted with two Voight peaks associated with the emission due to bright A neutral excitons ($A_0$) recombination and trions ($A_T$) recombination. (c) The schematic diagram for the energy levels in monolayer $WSe_2$ at the $K_+$ point of the Brillouin zone showing the direct electronic band-gap, binding energy for excitons and trions, and the recombination $A_0$ and $A_T$. According to reference [7], the electronic band gap is 2.0 eV, and the exciton binding energy is 0.37 eV. Hence, the neutral exciton emission from single-layer $WSe_2$ is expected at 1.65 eV, as observed here.

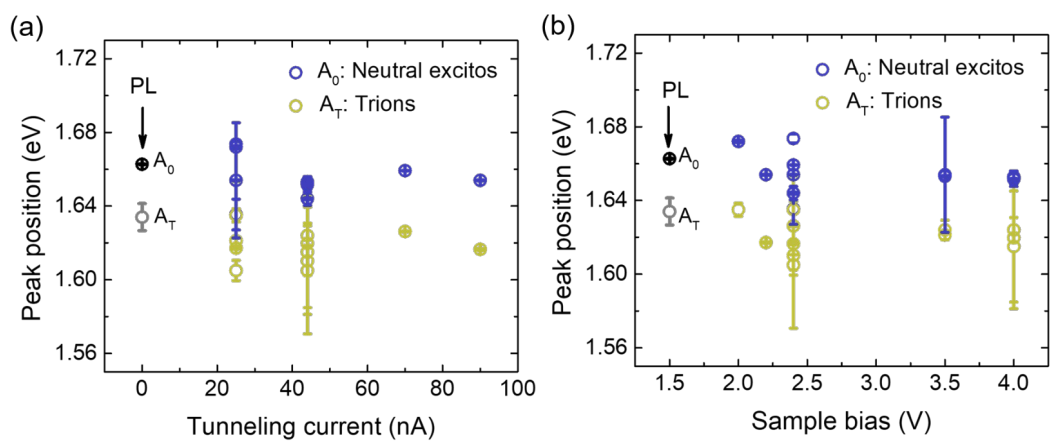

**Figure S8:** Position of the charged exciton (trion) and neutral exciton obtained from decomposing of STM-LE spectra. The spectra considered included those in Figure 2(c), 3(a), S6, and S7. In these cases, no plasmon contribution is observed, and all spectra can be decomposed as (and adjusted with) 2 peaks related to trion and neutral exciton emission. In (a) the position of the peaks is shown as a function of the tunneling current and in (b) as a function of the sample bias. In both cases, no systematic shift is observed. The small energy variation shown in these graphs is possibly related to the difference between each region on the monolayer $WSe_2$ that can be subjection to different stress states or have slightly different defect concentration. In (a) and in (b), the energy of both trion and neutral exciton obtained from low power PL (spectrum in Figure S7) is indicated for comparison. For the PL acquisition, there was no tunneling current or sample bias voltage.

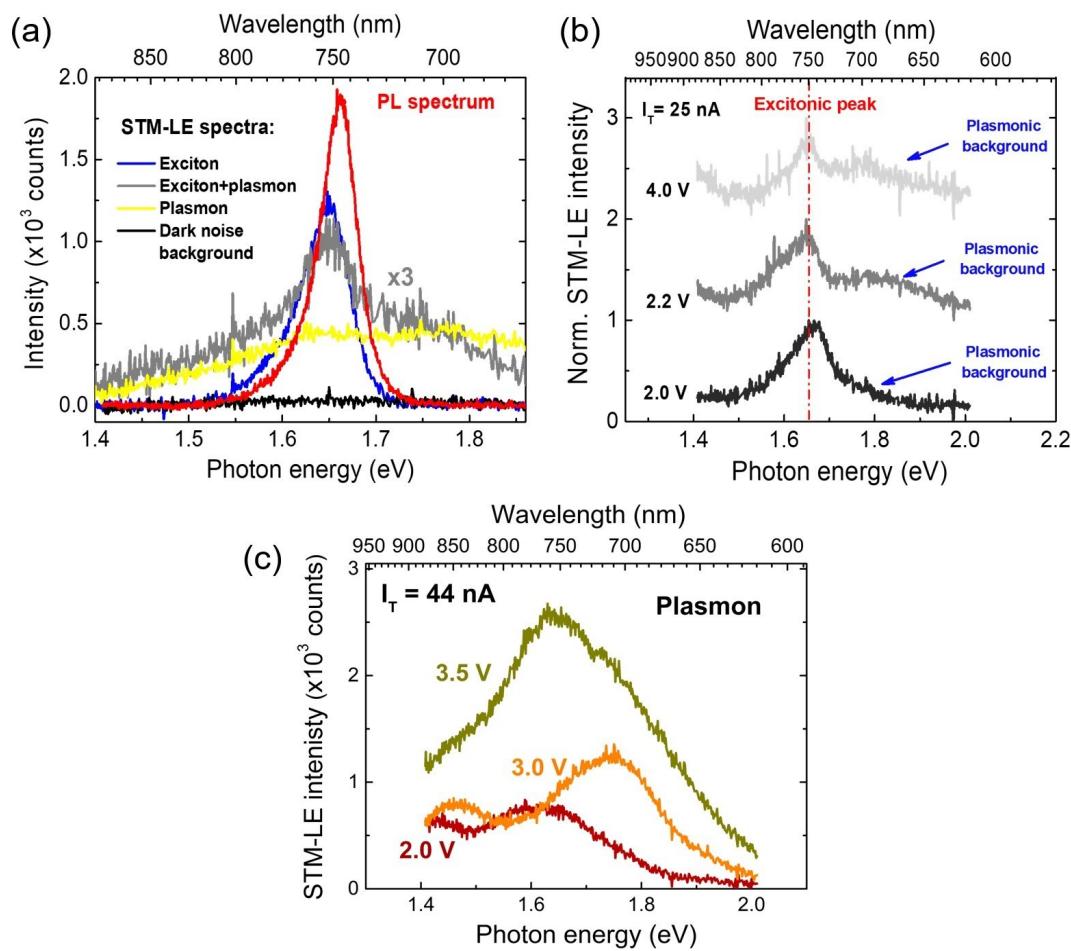

**Figure S9:** In (a), for comparison, the raw data of different spectra are shown, including the CCD dark noise, one PL spectrum, one STM-LE excitonic spectrum, one STM-LE with both plasmonic and excitonic emission and one STM-LE purely plasmonic. The STM-LE excitonic spectrum closely resembles the PL spectrum. In contrast, spectra with plasmonic contributions are readily spotted as much broader spectrally, covering the whole spectral range (as observed comparing them with the dark noise background). (b) STM-LE spectra of the simultaneous excitonic and plasmonic emission obtained with different tunneling parameters. (c) Pure plasmonic emission due to the gold metallic support for different sample bias voltage. The spectra obtained in (b) and (c) were recorded in the same sample region were high sample bias and high tunneling current was applied or in the same region were 'tip pulse' procedures were performed. A 'tip pulse' is a procedure in which the tip is put in contact with the sample (short circuit) for a brief moment to induce tip reconstruction.